\documentclass[aps,pre,twocolumn,superscriptaddress,showpacs,preprintnumbers]{revtex4-2}

\usepackage{amsmath,amssymb,amsfonts} 
\usepackage{algorithmic}
\usepackage{graphicx}
\usepackage{textcomp}
\usepackage{xcolor}
\usepackage{hyperref}
\usepackage{placeins}

\hypersetup{	
    colorlinks=true,
    linkcolor={blue!60!black},
    citecolor={blue!80!black},
    urlcolor={blue!80!black}
}

\begin{document}
	
\title{From Nodes to Edges: Edge-Based Laplacians for Brain Signal Processing}

\author{Andrea Santoro}
\thanks{\textcopyright{} 2025 IEEE. Personal use of this material is permitted. Permission from IEEE must be obtained for all other uses, in any current or future media, including reprinting/republishing this material for advertising or promotional purposes, creating new collective works, for resale or redistribution to servers or lists, or reuse of any copyrighted component of this work in other works. Accepted at the 33rd European Signal Processing Conference (EUSIPCO 2025). DOI: \href{https://doi.org/10.23919/EUSIPCO63237.2025.11226642}{10.23919/EUSIPCO63237.2025.11226642}}
\email{andrea.santoro@centai.eu}
\affiliation{CENTAI, C.so Re Inghilterra 3, 10138, Turin, Italy}

\author{Marco Nurisso}
\affiliation{CENTAI, C.so Re Inghilterra 3, 10138, Turin, Italy}
\affiliation{Dipartimento di Scienze Matematiche, Politecnico di Torino, Turin 10129, Italy}
\affiliation{SmartData@PoliTO Center, Politecnico di Torino, Turin 10129, Italy}

\author{Giovanni Petri}
\email{giovanni.petri@nulondon.ac.uk}

\affiliation{NPLab, Network Science Institute, Northeastern University London, London E1W 1LP, United Kingdom}
\affiliation{Department of Physics, Northeastern University, Boston, MA, USA}
\affiliation{CENTAI, C.so Re Inghilterra 3, 10138, Turin, Italy}

\begin{abstract}
Traditional graph signal processing (GSP) methods applied to brain networks focus on signals defined on the nodes.
Thus, they are unable to capture potentially important dynamics occurring on the edges. 
In this work, we adopt an edge-centric GSP approach to analyze edge signals constructed from 100 unrelated subjects of the Human Connectome Project. 
Specifically, we describe structural connectivity through the lens of the 1-dimensional Hodge Laplacian, processing signals defined on edges to capture co-fluctuation information between brain regions. 
We demonstrate that edge-based approaches achieve superior task decoding accuracy in static and dynamic scenarios compared to conventional node-based techniques, thereby unveiling unique aspects of brain functional organization. 
These findings underscore the promise of edge-focused GSP strategies for deepening our understanding of brain connectivity and functional dynamics.
\end{abstract}


\maketitle

\section{Introduction}
Graph Signal Processing (GSP) has emerged as a powerful framework that extends classical signal processing techniques to data defined on nonhomogeneous domains, such as graphs~\cite{sandryhaila2013discrete}. 
In many scientific domains, GSP provides a structured approach to analyze signals in the context of an underlying network architecture that is thought to support and generate the dynamics. 
More specifically, in neuroscience, diffusion-weighted imaging is typically used to construct structural connectomes $-$ the networks of white-matter pathways that connect distinct brain regions~\cite{atasoy2016human,atasoy2017connectomeharmonic}. 
Functional imaging techniques in turn capture dynamic signals that serve as direct or indirect proxies for neural activity~\cite{logothetis2002neural}. 
Among these, fMRI leverages the Blood Oxygenation Level-Dependent (BOLD) signal to track hemodynamic fluctuations that reflect underlying brain activity patterns~\cite{logothetis2003underpinnings,logothetis2004interpreting}.

Within the traditional GSP framework, each brain region is usually represented by a node in the graph given by the structural connectome, while the corresponding fMRI time series are treated as the signals defined on such nodes. 
By employing operators derived from the graph structure --- typically the graph Laplacian~\cite{belkin2003laplacian} --- one can perform a graph Fourier transform (GFT) to decompose these signals into their structural spectral patterns~\cite{shuman2013emerging}.
This decomposition not only respects the brain's anatomical backbone, but it also enables operations like graph filtering and randomization~\cite{pirondini2016spectral}.

Although GSP has proven effective for analyzing node-level signals, many neural processes hinge on interactions \textit{among nodes} --- defined at the edge and higher-order levels. 
Indeed, brain function emerges not only from the activity of isolated regions but also from the coordinated dynamics of pairwise connections and complex multi-way interactions among larger groups of regions. 
This insight has sparked a growing interest in higher-order descriptions~\cite{battiston2020networksa,battiston2021physicsa,bianconi2021higherorder}, which capture the intricate relationships invisible to conventional node-centric graph models.

To model these complex interactions, recent advances have extended signal analysis from nodes to higher-dimensional structures, typically using simplicial complexes~\cite{bianconi2021higherorder}.
In a simplicial complex, signals are not only defined on nodes (0-simplices), but can also be on edges (1-simplices), triangles (2-simplices), and higher-dimensional analogs. 
This enriched representation models complex relationships, such as the synchrony between pairs of regions, or the collective dynamics among groups~\cite{millan2025topology}. 
Topological Signal Processing (TSP) naturally extends the principles of GSP to data defined on higher-order structures~\cite{barbarossa2020topological,sardellitti2021topological,isufi2025topological}. 
For example, in brain network analysis, while node signals represent regional activation, signals on edges may capture co-activation or synchrony between pairs of regions, and signals defined on triangles can reveal interactions among three regions, thereby providing deeper insights into the coordinated activity underlying cognitive processes~\cite{santoro2023higherordera,santoro2024higherorder}.

Here, we introduce a novel TSP framework and apply it to the fMRI data of 100 unrelated subjects from the Human Connectome Project~\cite{vanessen2013wuminn}.
We outline methods to construct, filter, and project signals onto the eigenvectors of the Hodge Laplacian~\cite{lim2020hodge,anand2024hodge,roy2025hodge} --- a generalization of the graph Laplacian to simplicial complexes of any order. 
Using these techniques, we systematically analyze the temporal dynamics of higher-order interactions in neural circuits, with a focus on task decoding~\cite{barch2013function}. 
We find that edge-level descriptions provide better discrimination between tasks. 
Overall, by extending traditional graph signal processing to higher-order systems, our framework offers new insights into how complex neural connectivity patterns underpin brain function.

\section{Background}

In classical GSP, one considers a graph \( G = (V,E) \) with \( N \) vertices. 
Let \( A \in \mathbb{R}^{n \times n} \) be the adjacency matrix and \( D \in \mathbb{R}^{n \times n} \) be the diagonal degree matrix defined by \( D_{ii} = \sum_{j} A_{ij} \). 
The graph Laplacian is given by $L = D - A$. 
Alternatively, if we define the (oriented) node-edge incidence matrix \( B \in \mathbb{R}^{n \times m} \) (with \( m \) edges), then the graph Laplacian can be expressed as
\begin{equation}
    L = B B^T ,
\end{equation}
which admits an eigendecomposition of the form:
\begin{equation}
    L = U \Lambda U^T,
\end{equation}
where the columns of \( U \) are the eigenvectors and \( \Lambda \) is the diagonal matrix of the nonnegative eigenvalues. 
The \textit{graph Fourier transform} (GFT) of a signal \( x \in \mathbb{R}^{n} \) (defined on the nodes) is then given by 
\begin{equation}
    \hat{x} = U^T x,
\end{equation}
with the inverse transform \( x = U \hat{x} \). 
This framework allows us to perform operations such as filtering, denoising, and spectral analysis on signals that reside on the vertices of the graph~\cite{ortega2018graph,pirondini2016spectral}.

\subsection{Simplicial Complexes and the Hodge Laplacian}
Many complex systems exhibit interactions that extend beyond simple pairwise connections and therefore cannot be easily described as graphs. 
In these cases, it can be convenient to shift our focus away from nodes and consider signals defined on edges or on general groups of nodes.

TSP generalizes the classical GSP framework to deal with these higher-order signals by leveraging the topological theory of simplicial complexes: combinatorial objects built with nodes, edges, triangles, and their higher-dimensional analogs.

Formally, given a set of nodes $V$, a $k$-simplex $\sigma$ is a set of $k+1$ nodes in $V$, and a \emph{simplicial complex} $K$ is a set of simplices where any subset of a simplex $\sigma\in K$ is still a simplex in $K$.
We can think of $0$-simplices as nodes, $1$-simplices as edges, $2$-simplices as triangles, and so on.
Let \( n_k \) denote the number of \( k \)-simplices in \( \mathcal{K} \). 
A \( k \)-dimensional signal, usually named \emph{cochain}, is a function that associates a real number to every $k$-simplex. 
We can therefore think of the space of $k$-cochains as $\mathbb{R}^{n_k}$.
In the rest of the paper, we will mainly deal with edge signals, and thus we restrict the theory to the case $k=1$ (for a detailed formalism and generalization see~\cite{grady2010discrete,nurisso2024unified}).
The rationale for this choice is:
(i) We obtain significantly more information already for $k=1$, 
and (ii) the structural connectome is a natural candidate for the underlying structure, which can be seen as a 1-dimensional simplicial complex.

For each \( k \ge 1 \), the \emph{oriented boundary operator} \( B_k \in \mathbb{R}^{n_{k-1} \times n_k} \) connects the spaces of higher-order signals by mapping a \( k \)-simplex to a formal linear combination of its \((k-1)\)-faces $ B_k \in \mathbb{R}^{n_{k-1}\times n_k}$.
In our case, $B_1$ is equal to the node-edge incidence matrix $B_1 = B$, and $B_2$ describes incidence relations between edges and triangles.

The \emph{combinatorial Hodge Laplacian} of order \( 1 \) is defined as the sum
\begin{equation}
L_1 = L_1^\downarrow + L_1^\uparrow = B_1^T B_1 + B_{2} B_{2}^T,    
\end{equation}
where the first addendum describes interactions between edges and nodes, while the second describes interactions between edges and triangles.

As for the graph Laplacian, the edge-based Hodge Laplacian allows an eigendecomposition of the form:
\begin{equation}
L_1 = U_1 \Lambda_1 U_1^T,    
\end{equation}
where the columns of $U_1$ are the eigenvectors of $L_1$ and \( \Lambda_1 \) is the diagonal matrix of its eigenvalues. 
The \emph{topological Fourier transform} (TFT) of an edge-signal \( x_1 \in \mathbb{R}^{n_1} \) is then obtained by mapping it to the basis of the eigenvectors~\cite{barbarossa2020topological}
\[
\hat{x}_1 = U_1^T x_1,
\]
with the inverse transform given by \( x_1 = U_1 \hat{x}_1 \). 

\subsection{Hodge Decomposition}
A central result in TSP is the Hodge decomposition~\cite{isufi2025topological}, which states that any \( 1 \)-signal \( x \in \mathbb{R}^{n_1} \) can be uniquely decomposed as the sum of three orthogonal components:
\[
x = x_{\mathrm{grad}} + x_{\mathrm{curl}} + x_{\mathrm{harm}},
\]
where:

\begin{itemize}
    \item \textit{Gradient (Exact) Component}: \( x_{\mathrm{grad}} \) lies in the range of \( B_1^T \) and is expressed as
    \[
    x_{\mathrm{grad}} = B_1^T y, \quad \text{for some } y \in \mathbb{R}^{n_0}.
    \]
    This component captures variations that are induced by differences across adjacent nodes.
    
    \item \textit{Curl Component:} \( x_{\mathrm{curl}} \) lies in the range of \( B_{2} \) and is given by
    \[
    x_{\mathrm{curl}} = B_{2} z, \quad \text{for some } z \in \mathbb{R}^{n_{2}}.
    \]
    It contains the component of the signal that circulates around triangles.
    
    \item \textit{Harmonic Component:} \( x_{\mathrm{harm}} \) is the residual component that lies in the kernel of \( L_1 \):
    \[
    x_{\mathrm{harm}} = x - x_{\mathrm{grad}} - x_{\mathrm{curl}}.
    \]
    In practice, \( x_{\mathrm{harm}} \) is a linear composition of the harmonic eigenvectors of \( L_1 \), i.e., the ones associated with zero eigenvalues. 
    The topological significance of the harmonic space $\ker L_1$, stands in the fact that its dimension is equal to the number of 1-dimensional holes (cycles) in the simplicial complex~\cite{lim2020hodge}. 
    The harmonic eigenvectors that span it are also seen to be localized around the cycles~\cite{muhammad2006control}.
\end{itemize}

\medskip

In summary, traditional GSP focuses on signals defined on graph nodes using the graph Laplacian and its Fourier transform. 
The TSP framework extends these ideas to higher-order domains. 
In particular, by leveraging simplicial complexes, Hodge Laplacian, and Hodge decomposition, TSP allows for the analysis of signals on edges, triangles, and beyond. 
As we show in the next section, this generalized framework is particularly powerful for exploring complex interactions in brain function, in which multi-way relationships are critical to understanding neural dynamics~\cite{chialvo2010emergent}.

\begin{figure}
    \centering
    \includegraphics[width=1\linewidth]{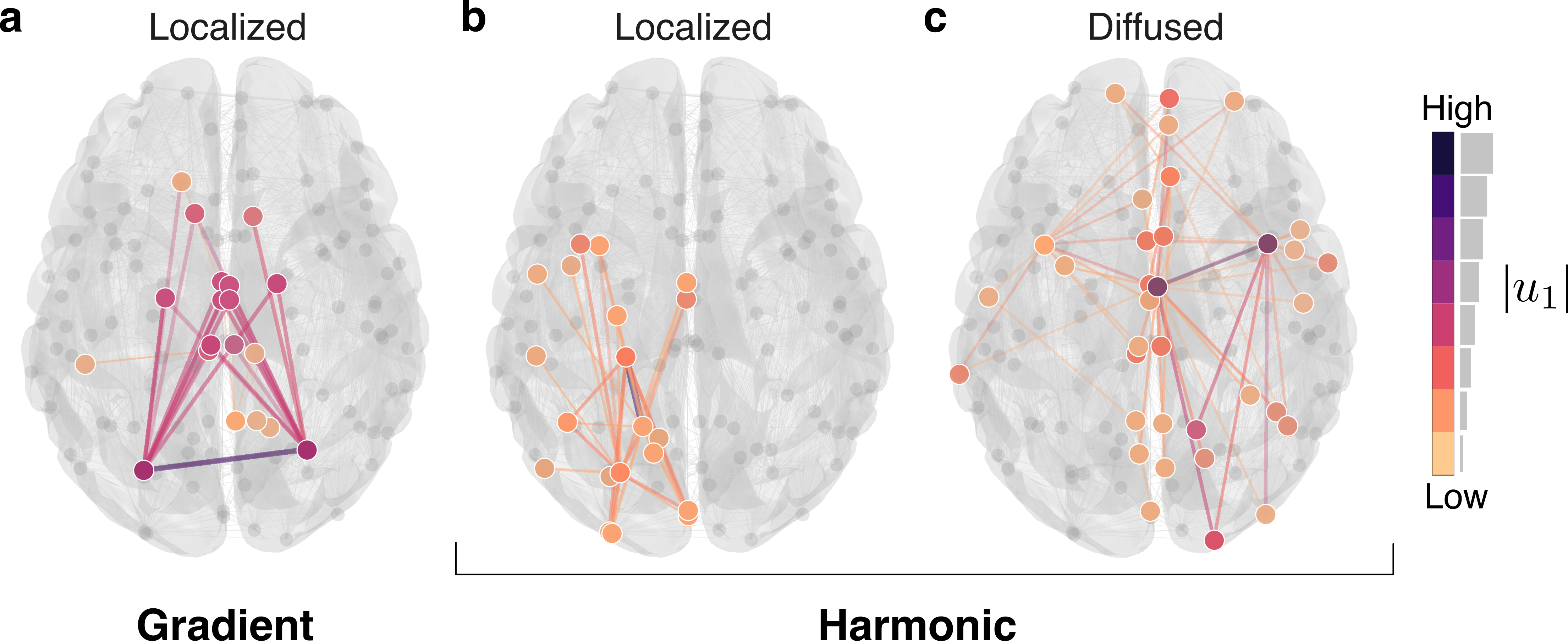}
    \caption{\textbf{Eigenmodes of the edge-based Hodge Laplacian $L_1^\downarrow$}. 
    The sparsest (according to the $\ell_1$ norm) gradient (\textbf{a}) and harmonic (\textbf{b}) eigenvectors of the edge-based Hodge Laplacian $L_1^\downarrow$. 
    \textbf{(c)} A non-sparse harmonic eigenvector. 
    The width and color of each edge represent the absolute value of its associated component in the eigenvector.}
    \label{fig:hodge_eigenmodes}
\end{figure}

\section{Application to Brain Signals}
\subsection{Methodology}
We analyze the structural and functional MRI data from resting state and seven different tasks of 100 unrelated healthy subjects from the Human Connectome Project (HCP)~\cite{vanessen2013wuminn,barch2013function}, considering the Schaefer 100~\cite{yeo2011organization} plus 19 subcortical regions (see~\cite{santoro2024higherorder} for preprocessing details). 
For these brain signals, we extract structure-function signatures --- both dynamically and statically --- using \textit{(i)} classical GSP approaches~\cite{ortega2018graph}, and \textit{(ii)} TSP approaches based on the edge-based Hodge Laplacian $L_1$ obtained from structural connectomes. 

Let $x_i(t)$ denote the time series at node $i$. 
We start by lifting the node time series to edge signals using two different approaches:

\textit{\textbf{i)}} The amplitude-based edge signal (also called co-fluctuation time series~\cite{faskowitz2020edgecentric}) between nodes $i$ and $j$ is obtained as:
\begin{equation}
    e_{ij}(t) = x_i(t)\, x_j(t).
\end{equation}
which quantifies the degree of co-activation between the corresponding brain regions.

\textit{\textbf{ii)}} Alternately, each node signal $x_i(t)$ is transformed via the Hilbert transform, denoted by $\mathcal{H}\left[x_i(t)\right]$, from which we retrieve the instantaneous phase $\theta_i(t)$.
The phases are then combined to form edge signals in the following way:
\begin{equation}\label{eq:cosine}
    e_{ij}(t) = f\big(\theta_i(t) - \theta_j(t)\big),
\end{equation}
where we chose $f$ as either the sine or cosine function to encode the edge-based phase synchronization between brain regions~\cite{lachaux1999measuringa,cabral2017cognitive}. 

We use the entire weighted structural connectome based on the number of white-matter fibers for the GSP approach, following recent methods by Preti et al.~\cite{preti2019decoupling} to examine both the coupled/decoupled signals and structural-decoupling index (SDI) signatures. 
For the TSP analysis, we first threshold the connectome to retain the top 20\% connections to balance sparsity and computational cost. 

We  compare two decomposition strategies: 
a first one in which we decompose only $L_1^\downarrow$ into harmonic and gradient components (Figure~\ref{fig:hodge_eigenmodes}), 
and a second one that fully leverages the TSP decomposition on $L_1$ to obtain the harmonic, gradient, and curl components. 
Notice that when constructing $L_1$, we consider the clique complex of the graph at order 2 --- treating each three-node clique as a $2$-simplex --- to account for higher-order interactions.

To compare the GSP and TSP approaches on fMRI data, we follow a methodology similar to~\cite{santoro2024higherorder}. 
That is, we first concatenate the initial 300 volumes of resting-state fMRI with data from seven task sessions --- excluding rest blocks and regressing out task paradigms --- to generate a unified fMRI recording. 
We then compute time–time correlation matrices from the filtered signals produced by the approaches, where each matrix entry $(i,j)$ represents the Pearson correlation between the temporal activations at time points $t_i$ and $t_j$. 

\begin{figure}[b!]
    \centering
    \includegraphics[width=\linewidth]{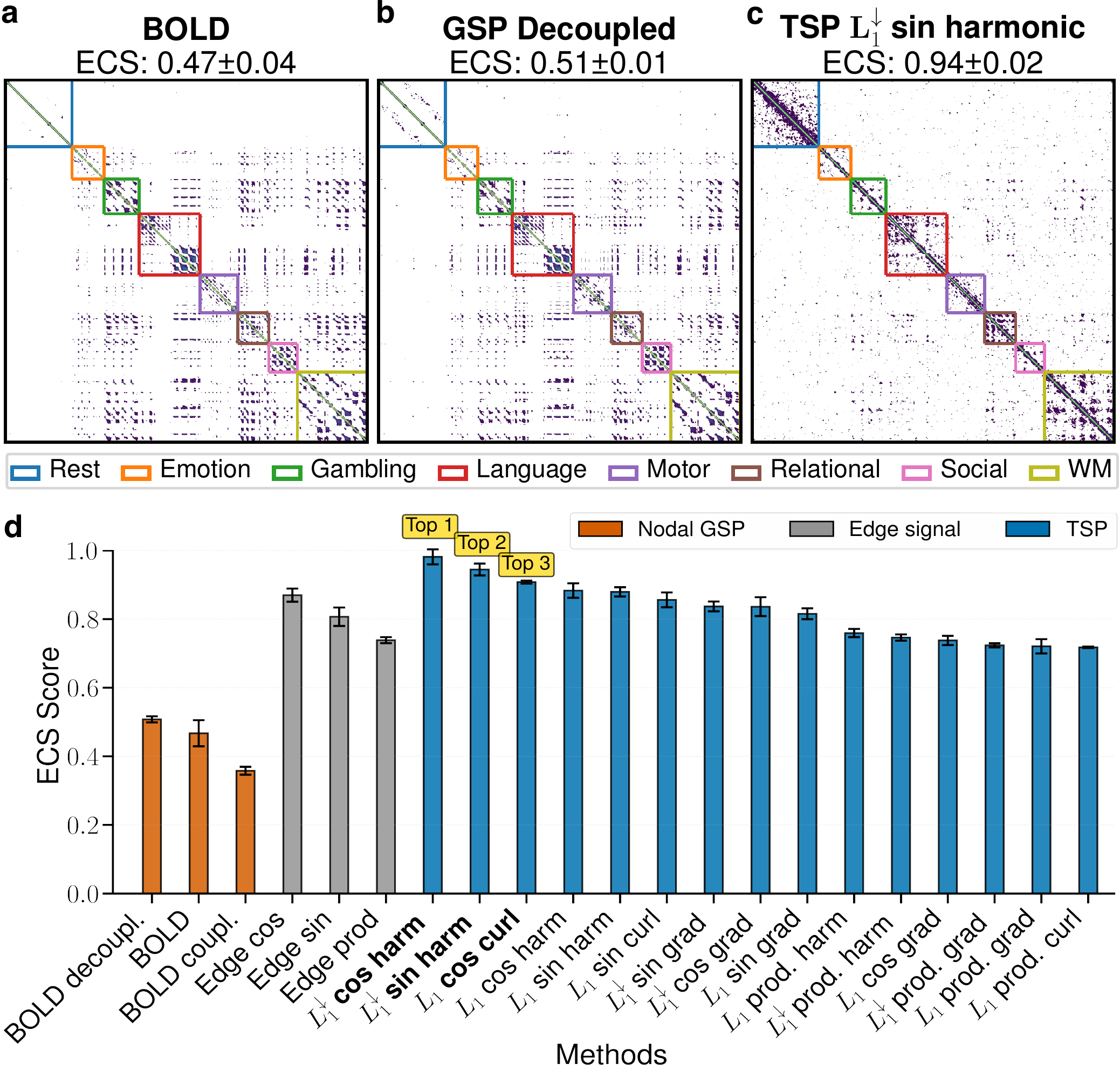}
    \caption{\textbf{Dynamic task decoding using GSP and TSP approaches}. 
    Filtered recurrence matrix (i.e. time-time correlations) for three representative measures, namely \textbf{(a)} BOLD signal, \textbf{(b)} GSP decoupled (high-frequency components) signals, and \textbf{(c)} TSP $L_1^{\downarrow}$ sin harmonic reconstructed signal. 
    In each case, the corresponding ECS metric (compared with the ground truth partition) is indicated. 
    \textbf{(d)} ECS bar plots for various GSP and TSP measures, with mean and standard deviation computed over 10 bootstrap samples, each using 80 out of the 100 subjects (considering LR and RL phase encoding, i.e. 160 subjects per sample). Notice that $L_1^\downarrow$ refers to the Laplacian constructed only with interactions between edges and nodes, whereas $L_1$ refers to the Laplacian accounting also for triangles as 2-simplex. Harm, grad, and curl correspond to different parts of the Hodge decomposition; prod, sin, and cos are the functions used to estimate the edge signals.}
    \label{fig:dyn_task_decoding}
\end{figure}

The correlation matrices, analogous to recurrence plots in dynamical systems, are binarized by thresholding at the 95th percentile (values below the threshold are set to zero, and those above are set to one). 
We then apply the Louvain algorithm~\cite{blondel2008fast} to extract community partitions, using consensus clustering (with 100 iterations) to address the algorithm's stochastic nature~\cite{lancichinetti2012consensus}. The effectiveness of these partitions in capturing the dynamical brain fluctuations task versus rest timing is then assessed using the element-centric similarity (ECS) measure~\cite{gates2019elementcentric}, which ranges from 0 for completely dissimilar partitions to 1 for perfectly corresponding ones.

\subsection{Results}
We report in Fig.~\ref{fig:dyn_task_decoding}\textbf{a-c} the results of our analysis, showing the recurrence plots of three representative measures, namely, the original BOLD signal, the coupled GSP-derived signal, and the $L_1^\downarrow$ harmonic component for the edge signal of Equation \eqref{eq:cosine} with $f=\sin$, each annotated with its ECS value. 
For completeness, we also provide in Fig.~\ref{fig:dyn_task_decoding}\textbf{d} the ECS measures for a range of other measures tested.

We now turn our attention to static task decoding, following the approach described in~\cite{griffa2022brain}. 
In this analysis, we employ a Support Vector Machine (SVM) to classify task-related states (rest plus seven tasks) based solely on node-wise measures of functional connectivity and structure-function coupling derived from GSP and TSP methods.  
That is, for each method, we construct nodal feature matrices of size  $N_{ROI} \times N_{BS}\, N_E\, N_S = 119 \times 1600$, where $N_{BS}$ represents the number of brain states (rest and tasks),  $N_E$ is the number of encoding conditions (2), and  $N_S$ is the number of subjects. 
As before, resting periods are excluded from the task, and task paradigms are regressed out from the fMRI time courses to minimize confounds from paradigm-imposed timings, aiming at keeping only differences due to the specific task-related states. 
For the GSP approach, we extract coupled, decoupled, and SDI nodal values (see~\cite{preti2019decoupling}), fixing a value $c = 30$ of spectral components to be common to all acquisitions and avoid task-biases that could affect the following classification. 
In contrast, for the TSP analysis, we first compute the $\ell_2$ norm over time of the filtered edge-level signal $x'_1 = \sqrt{\sum_{t=1}^T x_1(t)^2}$, and then project this measure onto the nodes using the boundary operator $B_1$. 
Formally, the nodal signal is defined as:
\begin{equation}
    s = |B_1| x'_1,
\end{equation}
where the $|B_1|$ is the element-wise absolute value of the boundary operator, to ignore the edge orientations.

\begin{figure}
    \centering
    \includegraphics[width=1\linewidth]{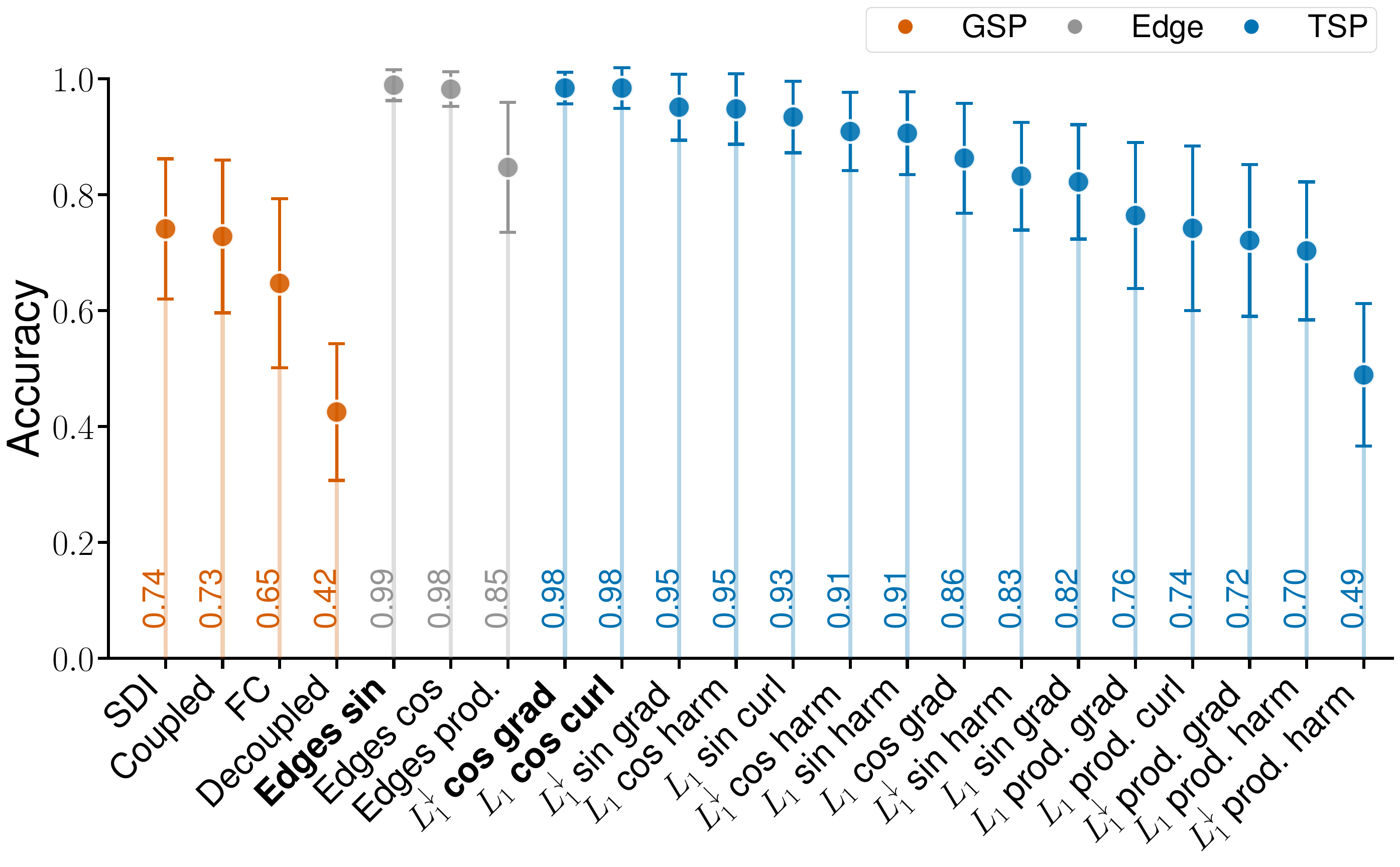}
    \caption{\textbf{Task decoding accuracy across methods.} 
    This figure presents the task decoding accuracies for node-wise functional connectivity and structure-function coupling measures derived from both GSP and TSP approaches. 
    Accuracies were computed using 100-fold leave-one-subject-out cross-validation with one-versus-one multiclass linear SVM classifiers. The top three performing methods are highlighted in bold. 
    Also in this case, $L_1^\downarrow$ refers to the Laplacian constructed only with interactions between edges and nodes, whereas $L_1$ refers to the Laplacian also accounting for triangles as 2-simplex. Harm, grad, and curl correspond to different parts of the Hodge decomposition; prod, sin, and cos are the functions used to estimate the edge signals.}
    \label{fig:static_task_decoding}
\end{figure}

Figure~\ref{fig:static_task_decoding} reports the accuracy of 100-fold leave-one-subject-out cross-validation and one-versus-one multiclass linear SVM for different GSP and TSP reconstructions. 
Interestingly, also in this case, we find that TSP-based measures outperform classical GSP approaches, remarking the importance of the dynamics of the connections (and thus of the Hodge decomposition), rather than that of the single nodal activations. 

\section{Discussion}
Understanding the intricate relationship between brain activity and its underlying anatomical connectivity has long posed significant challenges in neuroscience. Traditional approaches often employ large-scale neural population models that integrate structural connectivity to assess features observed in FC, such as modular organization and spatiotemporal dynamics~\cite{roberts2019metastable}. 
Graph signal processing offers a powerful approach, by treating the structural connectome as a graph, to decompose node-based signals (e.g., fMRI data) into various frequency components via the graph Fourier transform. 
While these methodologies have provided valuable insights into the distribution of neural activity across networks, their primary focus on node-level signals limits their capacity to capture more complex interactions.

In this work, we shifted our focus to an edge-centric perspective through Topological Signal Processing~\cite{isufi2025topological}, an extension of GSP that considers signals also defined on higher-order simplices, such as edges and triangles, within the network. 
By introducing a novel way to define temporal signals at the level of edges, we provide the first evidence that TSP methods outperform traditional GSP approaches in decoding cognitive tasks. 
Specifically, leveraging the decomposition of the Hodge Laplacian, we break down edge-level signals into gradient, curl, and harmonic components. 
Notably, our analysis reveals that filtering out the gradient component --- which reflects differences directly derived from node-level activity --- unmasks higher-order interactions encoded in the curl and harmonic subspaces.  
These refined edge-centric features significantly enhance task decoding performance, as confirmed by both dynamic analyses (using time–time correlation matrices) and static classification with Support Vector Machines.

Furthermore, we can map filtered edge signals back to the node level by using simple mathematical operations --- e.g. lifts and projections with boundary operators \cite{schaub2021signal} --- we can map filtered edge signals back to the node level. 
This process uncovers nuanced aspects of brain connectivity that are overlooked when focusing solely on node-based analyses.
Overall, our study underscores the potential of TSP-based approaches to capture and filter higher-order edge signals, thereby offering a more comprehensive understanding of inter-regional and intra-regional coupling variations over time or under different experimental conditions.
This innovative methodology not only enhances task decoding capabilities but also holds promise for providing fresh insights into connectivity alterations associated with neurological diseases.

\bibliography{biblio}

\end{document}